\documentclass[prl,twocolumn,showpacs,amsmath,amssymb,superscriptaddress]{revtex4}
\usepackage[dvips]{graphicx}% Include figure files--uses eps figures
\usepackage{graphics}% Include figure files
\usepackage{dcolumn}% Align table columns on decimal point
\usepackage{bm}% bold math

\newcommand{\kk}{{\bf k}}
\newcommand{\pp}{{\bf p}}

\begin{document}
\title{Ferromagnetic transition temperature enhancement
in (Ga,Mn)As semiconductor by carbon co-doping}
\author{T. Jungwirth}
\affiliation{University  of  Texas at  Austin,  Physics Department,  1
University   Station  C1600,   Austin  TX   78712-0264}
\affiliation{Institute of Physics  ASCR, Cukrovarnick\'a 10, 162 53
Praha 6, Czech Republic }
\author{J. Ma\v{s}ek}
\affiliation{Institute of Physics  ASCR, Na Slovance 2, 182 21
Praha 8, Czech Republic }
\author{Jairo Sinova}
\affiliation{University  of  Texas at  Austin,  Physics Department,  1
University Station C1600,  Austin TX 78712-0264}
\affiliation{Department of Physics, Texas A\&M University, College Station, TX 77843-4242}
\author{A.H. MacDonald} \affiliation{University of Texas at
Austin,  Physics Department,  1  University Station  C1600, Austin  TX
78712-0264}
\date{\today}
\begin{abstract}
We present a theoretical study of (Ga,Mn)(As,C) diluted
magnetic semiconductors with high C acceptor density
that combines insights from phenomenological model 
 and microscopic approaches. A
tight-binding coherent potential approximation is used to describe
the electronic structure in the presence of Mn$_{\rm Ga}$ and C$_{\rm
As}$ impurities. We find only a small effect of C on the
distribution and coherence of
electronic states close to the top of the valence band
and on the coupling between Mn moments,
even at doping levels of several per cent. These results justify applying
the  model of ferromagnetic Mn-Mn coupling mediated by
itinerant holes in the valence band also to  C
doped samples. The increase of ferromagnetic transition temperature
due to the presence of C acceptors is illustrated by
calculations that use the $\kk\cdot\pp$ Kohn-Luttinger
description of the GaAs valence band and assume systems where Mn local moment
and itinerant hole densities can
be varied independently.

\end{abstract}
\maketitle %it has to be after abstract and pacs

Prospects for new device
functionalities in all-semiconductor spin-electronic structures
rely on the realization of a ferromagnetic semiconductor
operating at room temperature. An important milestone in this
material research was the discovery five years ago \cite{ohno_science98}
of
ferromagnetism in  Mn-doped
GaAs \cite{firstGaMnAs}
with the Curie temperature $T_c=110$~K \cite{ohno_science98}.
In the Ga$_{1-x}$Mn$_x$As
diluted magnetic semiconductor (DMS) with $x\approx 1-10\%$, Mn
substituting for Ga provides a local
moment $S=5/2$ and a delocalized hole \cite{bookchapter,mm03}. Ferromagnetic
coupling between Mn moments is mediated by the itinerant holes
via a kinetic-exchange interaction \cite{dietl}. Recent
progress in low-temperature MBE
growth and post-growth annealing techniques has led to the
increase of the transition temperature in (Ga,Mn)As DMS's
by nearly 50~K \cite{anneal}.
This success is attributed to a smaller concentration
of carrier and moment compensating defects, especially interstitial Mn,
in the optimally annealed samples \cite{Mn-inter}. The latest
observations are consistent with
an approximately linear dependence of $T_c$ on $x$ and
hole Fermi wavevector predicted by theory
\cite{jungwirth_prb99,dietl_science00,jungwirth_prb02}.

In this paper we address theoretically the possibility of increasing
$T_c$ in (Ga,Mn)As DMS's by a non-magnetic acceptor co-doping, namely
by introducing substitutional C$_{\rm As}$ impurities.
Our work is partly motivated
by a recent experimental observation of
 a marked enhancement of the  Curie temperature
in Mn-implanted GaAs:C samples compared to the Mn-implanted undoped
GaAs layers \cite{park_tbp}.
After briefly discussing the different effects of C$_{\rm As}$ and
Be$_{\rm Ga}$ \cite{Be-Furdyna}
acceptors on substitutional Mn
incorporation
in (Ga,Mn)As DMS's, we focus on two key issues related to 
high-density carbon doping: (i) The effect of C
impurities on the density of states in the semiconductor valence band is
assessed using the tight-binding/coherent-potential-approximation (TB/CPA)
description of the disordered semiconductor. (ii) The strength
of the  hole-mediated ferromagnetic Mn-Mn coupling is compared for systems
with and without C co-doping using the TB/CPA results and $T_c$ is then
estimated from a model that combines the $\kk\cdot\pp$ Kohn-Luttinger
description of the GaAs valence bands and a mean-field treatment of the
kinetic-exchange
coupling between Mn local moments and  the band-holes.

An important issue for hole co-doping in (Ga,Mn)As DMS's is the
change of the substitutional Mn$_{\rm Ga}$ formation energy caused
by the presence of  additional non-magnetic acceptors. Although
the formation energy is, strictly speaking, an equilibrium
characteristic we assume that the basic features of  its
compositional dependence are reflected also in the dynamics of the
non-equilibrium growth of the co-doped materials. A systematic
ab-inito study of the formation energies is beyond the scope of
this short paper and will be discussed elsewhere \cite{masek_tbp}.
Here we present a qualitative analysis based on the TB/CPA
calculations, described in more detail below. Our results suggest
that Be acceptors substituting for the same element as Mn, i.e.
for Ga, lead to a strong enhancement of the Mn$_{\rm Ga}$
formation energy. At the same time, the formation energy of
interstitial Mn is suppressed and, hence, an increasing fraction
of Mn is incorporated in the form of interstitial donors or
electrically neutral MnAs or Mn clusters. This scenario, which
leads to a decrease rather than an increase of $T_c$, has recently
been established by extensive experimental studies of Be co-doped
(Ga,Mn)As samples \cite{Be-Furdyna}.

The TB/CPA calculations indicate that C acceptors lead to a much
weaker increase of the Mn$_{\rm Ga}$ formation energy. We surmise
that this property stems from  the anomalous nature of C$_{\rm
As}$ which acts as an acceptor, yet has a larger Pauling's
electro-negativity than As. The difference between the effects of
Be$_{\rm Ga}$ and C$_{\rm As}$ impurities on Mn$_{\rm Ga}$
incorporation is further enhanced at high doping levels where the
competition of Be and Mn for the same lattice
site starts to play a role.

The above qualitative analysis of formation energies suggests
that  C$_{\rm As}$ co-doping
is favorable for achieving high Curie temperatures.
Unlike typical
acceptors, however, C has a very different atomic size and, as already
mentioned,
a large electro-negativity compared to the atom it substitutes for. The
crucial question, addressed in the following paragraphs, is then
to what extent doping by several per cent of C$_{\rm As}$ changes the
semiconductor band structure and whether the model of shallow acceptor
carrier-induced ferromagnetism still applies in (Ga,Mn)(As,C).

Our calculations of the electronic structure of GaAs in the
presence of  Mn$_{\rm Ga}$ and C$_{\rm
As}$ impurities are done using the tight-binding  version of
the coherent potential approximation  (see e.g.
Ref.~\cite{Masek87}). The CPA, in contrast to supercell calculations,
is well-suited for the mixed crystals with low concentrations of
randomly distributed impurities. It provides estimates of 
a configurationally
averaged density of states (DOS) and related quantities.
The configurational averaging restores full translational symmetry
of the lattice and makes it possible to decompose the DOS into
contributions from specific points in the Brillouin zone. The
spectral density $A(k,E)$ then includes  
a detailed information
about the dispersion of the electronic states in the reciprocal
space. In addition, the width of the peaks of $A(k,E)$ defines the
scattering rate of the band electrons in various parts of the
Brillouin zone due to the impurities.

The parameterization of the TB Hamiltonian provides a
correct band gap for the pure GaAs crystal \cite{Talwar82} and an
appropriate exchange splitting of the Mn d--states.
Local changes of the crystal potential at both Mn and C impurities,
represented by shifted atomic levels, are estimated using
Ref.~\cite{Harrison_book}.
Long-range tails of the impurity potentials, which become less
important with increasing level of doping, are neglected. (Note, that the
Thomas-Fermi screening length is only 3-5~$\AA$ for typical carrier
densities \cite{jungwirth_apl02}, i.e., comparable to the lattice constant.)
Also
lattice relaxation effects are neglected within the CPA. This is
well justified for Mn$_{\rm Ga}$ impurities
\cite{mirbt_jpcm02,masek_prb03} but becomes a more important issue
in the case of
C$_{\rm As}$ impurities. Previous density-functional studies
found a -0.6\% relative change of the GaAs lattice constant in a
64-atom super-cell with a single C$_{\rm As}$ impurity
\cite{latham_prb01}. Although the lattice relaxation around
C$_{\rm As}$ may change our results quantitatively, we expect that
 the dominant effects on the band structure and on the spectral broadening
arise from the different atomic levels of C compared to As which
is readily accounted for in the TB scheme.

\begin{figure}[h]
\includegraphics[width=2.2in,angle=-90]{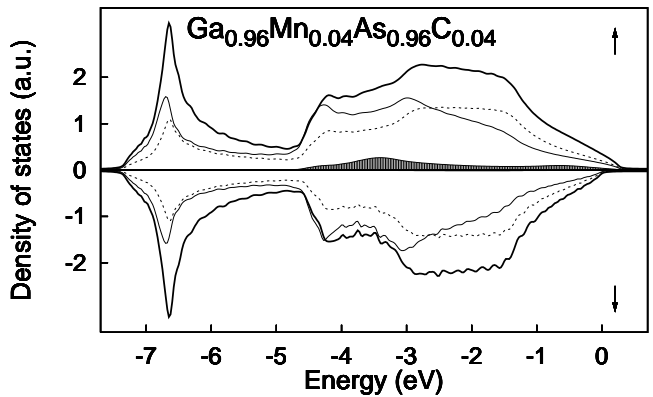}
\caption{TB/CPA density of states in
Ga$_{1-x}$Mn$_x$As$_{1-y}$C$_y$ DMS with $x=y=4$\%. Total DOS
(thick full line), local DOS on host As (thin dashed line) and
impurity C$_{\rm As}$ (thin full line) atoms, and the DOS of Mn
d-states are plotted as a function of energy, measured from the Fermi
level. } \label{dos}
\end{figure}

In Fig.~\ref{dos} we plot  the spin-polarized DOS in
Ga$_{1-x}$Mn$_x$As$_{1-y}$C$_y$ with $x=y=4$\% together with local
DOS on host (As) and impurity (C$_{\rm As}$) atoms. Despite the
remarkable difference between the atomic levels of As and C,
$\Delta
\varepsilon_{p} \approx$ -1 eV,  the local DOS on C$_{\rm As}$
sites (thin full line) near the valence band edge does not differ
much from the local DOS on the As sites (dashed line). (A still larger
full width of the valence band than $\Delta
\varepsilon_{p}$ may partly explain this result.) Similarly,
the total DOS (thick full line) is only weakly affected by
the presence of C in this spectral region. In particular, the
spin-splitting 0.26~eV of the valence band edge, directly related
to the kinetic-exchange parameter $J_{pd}$, is nearly the same in
systems with and without C$_{\rm As}$ impurities. The shaded
region in Fig.~\ref{dos} shows the DOS of Mn d-states peaked near
the energy of -4eV which is consistent with results in C free
(Ga,Mn)As.

\begin{figure}[h]
\includegraphics[width=2.2in,angle=-90]{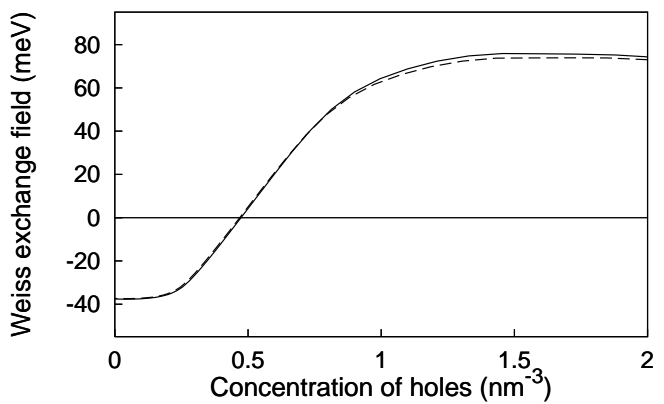}
\caption{Energy cost of flipping one Mn$_{\rm Ga}$ moment, obtained
from the TB/CPA spectra, is plotted as a function of the hole density for
$x=y=4$\% (solid
line), and $x=4$\% and $y=0$ (dashed line)}
\label{heff}
\end{figure}

It is important to point out that the substitution of C also has a
very small effect on the line-width of $A(k,E)$ for the states
close to top of the valence band, i.e., electron scattering on
C impurities does not disturb substantially the coherence of the
Bloch states. This, together with an unchanged value of $J_{pd}$,
implies that the additional disorder due to the co-doping with C
should not have any marked effect on the carrier mediated coupling
between Mn moments. We check this more explicitly using the
compatibility of the CPA with the Weiss mean-field theory. The
strength of the  Mn-Mn coupling is characterized by the energy
cost of flipping one Mn$_{\rm Ga}$ moment, which can be calculated
for a given chemical composition \cite{masek_ssc91}. In
Fig.~\ref{heff}, we plot this quantity as a function of the hole
density, $p$, for $x=y=4$\% (solid line), and $x=4$\% and $y=0$
(dashed line) samples.  The hole density is varied in the
calculations indepenedently of $x$ and $y$, i.e., we assume
implicitly a compensation whenever $p$ is smaller than the total density
of Mn$_{\rm Ga}$ and C$_{\rm As}$ acceptors.
The curves are nearly identical for systems with
and without C, as anticipated above. This means that, for a given
hole concentration, the exchange coupling is quite insensitive to
the large amount of the additional C$_{\rm As}$ defects. We
conclude our TB/CPA considerations by stating that, as in the
Mn$_{\rm Ga}$ acceptor case, the main effect of C$_{\rm As}$ doping
is a downward shift of Fermi energy with respect to the band
edge, i.e., an increase of the number of holes in the valence
band. In the following paragraphs we discuss prospects for
ferromagnetic transition temperature enhancement by adding extra
holes into the DMS valence band.

We use a model in which the itinerant holes are described by the
GaAs host bands and  the coupling to the local moments by a
phenomenological constant $J_{pd}$ \cite{dietl,ohno_jmmm99}. This
theory has been successful in describing semiquantitatively many
non-trivial thermodynamic and transport properties of (Ga,Mn)As
DMS's
\cite{dietl_science00,jungwirth_prb02,aniso,domains,jungwirth_prl02,jungwirth_0302060,sinova_prb02}. In the simplest,
virtual-crystal mean-field version of the model, that assumes
ferromagnetic indirect coupling between Mn ions, the energy to
flip the Mn$_{\rm Ga}$ moment is proportional to the effective
field $H_{eff}=J_{pd}\langle s\rangle$. Here $\langle s\rangle$ is
the mean spin-polarization density of the itinerant holes which
increases with the hole density. This result is consistent with
our TB/CPA calculations, shown in Fig.~\ref{heff}, for hole
densities not too much larger than the Mn$_{\rm Ga}$ density.
Recall that the concentration of the substitutional Mn ions is
given by $N_{Mn}=4x/a^3_{lc}$ where $a_{lc}$ is the GaAs lattice
constant ($N_{Mn}=0.88$~nm$^{-3}$ for $x=4$\%, e.g.). For
$p>N_{Mn}$,  RKKY oscillations of the Mn-Mn coupling
\cite{dietl} start to play a role and the increasing number
of antiferromagnetically coupled Mn moments leads to a saturation
or even to a suppression of $T_c$, as seen in Fig.~\ref{heff} for
$p>1.3$~nm$^{-3}$. The mean-field theory that allows only for
collinear ferromagnetic states is therefore
likely to break down in this high hole-density region. Note also
that for high $N_{Mn}$ and low $p$, the direct antiferromagnetic
Mn-Mn interaction takes over, as suggested by negative Weiss
exchange-field values in Fig.\ref{heff}, which sets another limit
on the validity of the simple mean-field model. This low
hole-density region is, however, not important for the high-$T_c$
(Ga,Mn)(As,C) DMS's we focus on in this paper.

We will use now the mean-field model to
estimate the Curie temperature as a function of the Mn$_{\rm Ga}$ local moment
density and of the density of itinerant holes that can be varied
independently. We emphasize that hole doping due to
C is relevant also for systems with
$p<N_{Mn}$ since most of the experimental (Ga,Mn)As samples
show some level of compensation, usually caused by interstitial
Mn defects.
In the
calculations, the
GaAs host band-structure is obtained from the $\kk\cdot\pp$, Kohn-Luttinger
model \cite{aniso}. We
neglect the $\sim 10-20$\% suppression of $T_c$ due to
spin-wave fluctuations since the effect is
compensated, to a large extent,
by hole-hole exchange enhancement of $T_c$
\cite{jungwirth_prb02}, also
neglected in the present calculations.
Details of the model are described elsewhere \cite{jungwirth_prb99,
bookchapter,aniso,jungwirth_prb02}; here we recall only the general $T_c$
expression we use \cite{dietl,jungwirth_prb99,jungwirth_prb02}:
\begin{equation}
k_BT_c=\frac{N_{Mn}S(S+1)}{3}\frac{J^2_{pd}\chi}{(g\mu_B)^2}\; ,
\end{equation}
where $\chi$ is the band-hole magnetic susceptibility which is roughly
proportional to $p^{1/3}$.

\begin{figure}[h]
\includegraphics[width=3.8in]{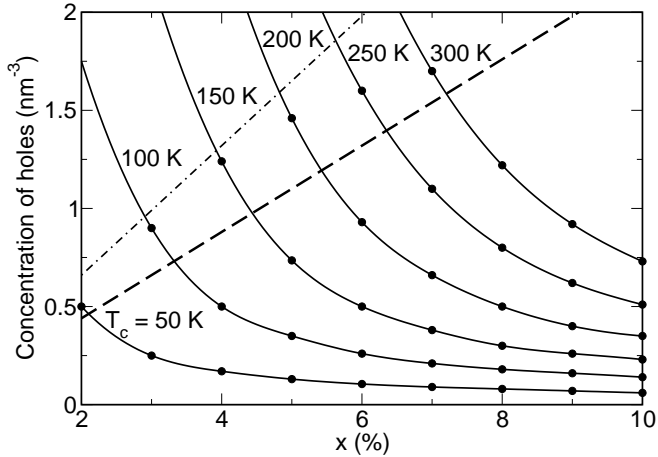}
\caption{Mean-field, $\kk\cdot\pp$ model calculations of constant $T_c$
curves are plotted in the Mn$_{\rm Ga}$ doping ($x$) -- hole density ($p$)
diagram. Dashed
line corresponds to $p=N_{Mn}=4x/a^3_{lc}$ and the dotted-dashed line indicates
approximately hole densities at which the mean-field model is expected to
break down due to RKKY oscillations of the Mn-Mn indirect coupling.}
\label{tc}
\end{figure}

Fig.~\ref{tc} shows constant-$T_c$ curves calculated for critical
temperatures ranging from 50~K to room-temperature. The dashed line
corresponds to $p=N_{Mn}$ and also indicates approximately
the onset
of
RKKY oscillation effects on $T_c$.
Note that the simple mean-field theory used to calculate the constant-$T_c$
curves is expected  to break down above the dotted-dashed line where
the large number of antiferromagnetically RKKY-coupled Mn moments
might lead to a decrease rather than increase of $T_c$ with increasing hole
density.

Fig.\ref{tc} suggests that a substantial enhancement of the ferromagnetic
transition temperature may be expected in hole co-doped samples with high
Mn moment concentration.
Assuming, e.g., $x=10$\%, an increase of the
hole density from  $p\approx 0.1$~nm$^{-3}$ to
$p\approx 0.7$~nm$^{-3}$ leads to an
increase of the theoretical $T_c$ from 50~K to 300~K.
For smaller $x$, a larger $\Delta p$ is needed to enhance $T_c$ by the
same amount.
Recent experiments have demonstrated that good quality (Ga,Mn)As DMS's
can be grown with $x$ reaching 8\% \cite{anneal}.
Based on the curves in
Fig.~\ref{tc} and our TB/CPA results we conclude,
that (Ga,Mn)As DMS's co-doped with
several per cent of C should not be overlooked among potential candidates
for a room-temperature ferromagnetic semiconductor.

\begin{acknowledgments}
We acknowledge fruitful discussions with Cammy Abernathy, Miroslav Cukr,
V\'{\i}t Nov\'ak, and Dan Park.
 This work was supported in part by the Welch Foundation,
DOE under grant DE-FG03-02ER45958, the Grant Agency of ASCR under
grant A1010214, and the Grant Agency of the Czech Republic under
grant 202/02/0912.
\end{acknowledgments}

%\bibliography{dms}

\begin{references}

\bibitem{ohno_science98} H. Ohno, Science {\bf 281}, 951 (1998).

\bibitem{firstGaMnAs}
H. Ohno, A. Shen, F. Matsukura, A. Oiwa, A. Endo, S. Katsumoto, and Y. Iye,
Appl. Phys. Lett. {\bf 69}, 363 (1996);
J. De Boeck, R. Oesterholt, H. Bender, A. Van Esch, C. Bruynseraede,
C. Van Hoof, and G. Borghs,  J. Magn. Magn. Mater. 156, 148 (1996).

\bibitem{bookchapter}
J. K\"{o}nig, J. Schliemann, T. Jungwirth,
and A.H. MacDonald,
in
{\em Electronic Structure and Magnetism of Complex Materials},
edited by D.J. Singh and D.A. Papaconstantopoulos (Springer Verlag, Berlin
2003) and citations therein.

\bibitem{mm03}
T.C. Schulthess, Bull. Am. Phys. Soc. 2003, March Meeting, K30-1;
P.H. Dederichs, K. Sato, H. Katayama-Yoshida, J. Kudrnovsk\'y,
{\em ibid}, S24-5.

\bibitem{dietl}
T. Dietl, "Diluted Magnetic Semiconductors" in Handbook of
Semiconductors, Vol. 3B", (North-Holland, New York), (1994); T.
Dietl, A. Haury and Y. Merle d'Aubign\'e, Phys. Rev. B {\bf 55},
R3347 (1997).

\bibitem{anneal}
K.W. Edmonds, K.Y. Wang, R.P. Campion, A.C. Neumann, N.R.S. Farley,
B.L. Gallagher, and C.T. Foxon,
Appl. Phys. Lett. 81, 4991 (2002);
K. C. Ku, S. J. Potashnik, R. F. Wang, M. J. Seong, E. Johnston-Halperin,
R. C. Meyers, S. H. Chun, A. Mascarenhas, A. C. Gossard, D. D. Awschalom,
P. Schiffer, and N. Samarth,
Appl. Phys. Lett. {\bf 82}, 2302 (2002);
K.W. Edmonds, K.Y. Wang, R.P. Campion, N.R.S. Farley,
B.L. Gallagher, C.T. Foxon, and M. Sawicki, to be published.

\bibitem{Mn-inter}
K. M. Yu, W. Walukiewicz, T. Wojtowicz, I. Kuryliszyn, X. Liu, Y. Sasaki,
and J. K. Furdyna, Phys. Rev. B {\bf 65}, 201303 (2002).

\bibitem{jungwirth_prb99} T.  Jungwirth,
W.A. Atkinson,  B.H. Lee,  and A.H. MacDonald,  Phys. Rev. B  {\bf 59}, 9818
(1999).

\bibitem{dietl_science00} T.  Dietl, H. Ohno, F.  Matsukura, J. Cibert,
and D. Ferrand, Science {\bf 287}, 1019 (2000).

\bibitem{jungwirth_prb02} T.  Jungwirth,  J.  K\"{o}nig, J.  Sinova, J.
Ku\v{c}era, and A.H.  MacDonald, Phys.  Rev.  B {\bf 66}, 012402 (2002).

\bibitem{park_tbp}
Y.D. Park, J.D. Lim, K.S. Suh, S.B. Shim, J.S. Lee, C.R. Abernathy,
S.J. Pearton, Y.S. Kim, Z.G. Khim, and R.G. Wilson, to be published.

\bibitem{Be-Furdyna}
K. M. Yu, W. Walukiewicz, T. Wojtowicz, W.L. Lim, X. Liu, U. Bindley, M. Dobrowolska, and J. K. Furdyna,
cond-mat/0303217.

\bibitem{masek_tbp}
J. Ma\v{s}ek, I. Turek, V. Drchal, J. Kudrnovsk\'{y}, and F.
M\'aca, to be published.


\bibitem{Masek87}
J.~Ma\v{s}ek, B.~Velick\'{y}, V.~Jani\v{s}, J. Phys. C {\bf 20},
59 (1987).

\bibitem{Talwar82}
D.~N.~Talwar, C.~S.~Ting, Phys. Rev. B {\bf 25}, 2660 (1982).

\bibitem{Harrison_book}
W. Harrison, {\it Electronic Structure and the Properties of
Solid} (Freeman, San Francisco, 1980).

\bibitem{jungwirth_apl02}
T. Jungwirth, M. Abolfath, Jairo Sinova, J. Ku\v{c}era, and A.H. MacDonald,
Appl. Phys. Lett. {\bf 81}, 4029 (2002);
M.P. L\'opez-Sancho and L. Brey, cond-mat/0302237.

\bibitem{mirbt_jpcm02}
S. Mirbt, B. Sanyal, and P. Mphm, J. Phys.: Condens. Matter {\bf
14}, 30~130 (2002).

\bibitem{masek_prb03}
J. Ma\v{s}ek, J. Kudrnovsk\'{y}, and F. M\'aca, Phys. Rev. B
{\bf67}, 153203 (2003).

\bibitem{latham_prb01}
C.D. Latham, R. Jones, S. \"{O}berg, and P.R. Briddon,
Phys. Rev. B {\bf 63}, 155202 (2001).

\bibitem{masek_ssc91}
J.~Ma\v{s}ek, Solid State Comm. {\bf 78}, 351 (1991).

\bibitem{ohno_jmmm99}
H. Ohno, J. Magn. Magn. Mater, {\bf 200}, 110 (1999);
J. Okabayashi, A. Kimura, O. Rader, T. Mizokawa, A. Fujimori,
T. Hayashi, and M. Tanaka, Phys. Rev. B {\bf 58}, R4211 (1998).

\bibitem{aniso}
T. Dietl, H. Ohno, F. Matsukura,  Phys. Rev. B {\bf 63}, 195205 (2001);
M. Abolfath, T. Jungwirth, J. Brum, and A.H. MacDonald,
Phys. Rev. B {\bf 63}, 054418 (2001).

%\bibitem{konig_prb01}
%J. K\"{o}nig, T. Jungwirth, and A.H. MacDonald,
%Phys. Rev. B {\bf 64}, 184423 (2001).

%\bibitem{stiffness}
%S.T.B. Goennenwein, T. Graf, T. Wassner, M.S. Brandt,
%M. Stutzmann, J. B. Philipp, R. Gross, M. Krieger, K. Z\"{u}rn,
%P. Ziemann, A. Koeder, S. Frank, W. Schoch, and A. Waag,
%Appl. Phys. Lett. {\bf 82}, 730 (2003).

\bibitem{domains}
T. Dietl, J. K\"{o}nig, and A. H. MacDonald,
Phys. Rev. B {\bf 64}, 241201 (2001).

\bibitem{jungwirth_prl02}
T. Jungwirth, Q. Niu, and A.H. MacDonald, Phys. Rev. Lett.
{\bf 88}, 207208 (2002).

\bibitem{jungwirth_0302060}
T. Jungwirth, Jairo Sinova, K.Y. Wang, K. W. Edmonds,
R.P. Campion, B.L. Gallagher, C.T. Foxon, Q. Niu, A.H. MAcDonald,
preprint cond-mat/0302060.

\bibitem{sinova_prb02}
Jairo Sinova, T. Jungwirth, S.- R. Eric Yang, J. Kucera, and A.H. MacDonald,
Phys. Rev. B {\bf 66}, 041202 (2002).


\end{references}

%%%%the references below have been generated by bibtex:
\end{document}